\documentclass[amsmath,amssymb,aps,pre,reprint,superscriptaddress,citeautoscript]{revtex4-1}

\usepackage{graphicx}
\usepackage{amsmath,array,float,subfigure}
\usepackage[makeroom]{cancel}
\usepackage[dvipsnames,svgnames,x11names,graphicx,array,empheq,tikz,amsthm,amssymb,multirow,amsfonts,ulem]{xcolor}
\usepackage[markup=underlined]{changes}

\begin{document}
\preprint{APS/123-QED}

\title{Large amplitude dust-acoustic solitary waves and double layers in nonthermal warm complex plasmas}
\author{N. Alam}
\affiliation{Department of Physics, Jahangirnagar University, Savar, Dhaka-1342, Bangladesh}
\email{antoralam55@gmail.com}
\author{A. Mannan}
\affiliation{Department of Physics, Jahangirnagar University, Savar, Dhaka-1342, Bangladesh}
\affiliation{Laboratori Nazionali di Frascati, INFN, Via Enrico Fermi 54, 00044, Frascati, RM, Italy}
\author{A A Mamun}
\affiliation{Department of Physics, Jahangirnagar University, Savar, Dhaka-1342, Bangladesh}

\date{\today}

\begin{abstract}
Using a Sagdeev pseudopotential approach where the nonlinear structures are stationary in a comoving frame, the arbitrary or large amplitude dust-acoustic solitary waves and double layers have been studied in dusty plasmas containing warm positively charged dust and nonthermal distributed electrons and ions. Depending on the values of the critical Mach number, which varies with the plasma parameter, both supersonic and subsonic dust-acoustic solitary waves are found. It is found that our plasma system under consideration supports both positive and negative supersonic solitary waves, and only positive subsonic solitary waves and negative double layers. The parametric regimes for the existence of subsonic and supersonic dust-acoustic waves and how the polarity of solitary waves changes with plasma parameters are shown. It is observed that the solitary waves and double layers solution exist at the values of Mach number around its critical Mach number. The basic properties (amplitude, width, speed, etc.) of the solitary pulses and double layers are significantly modified by the plasma parameters (viz. ion to positive dust number density ratio, ion to electron temperature ratio, nonthermal parameter, positive dust temperature to ion temperature ratio, etc.). The applications of our present work in space environments (viz. cometary tails, Earth's mesosphere, Jupiter's magnetosphere, etc.) and laboratory devices, where nonthermal ions and electrons species along with positively charged dust species have been observed, are briefly discussed.
\end{abstract}

\maketitle
\section{Introduction}
\label{1sec:Introduction}
The physics of dusty plasmas has recently attracted more attention because of the large number of dust particles in our universe and their significance in understanding numerous collective processes in astrophysical and space environments \cite{Shukla2001,Mendis1994,Goertz1989,Verheest2000}. The most recent discoveries in the study of such complex plasma systems concern the remarkable capacity of small particles to emerge from molecular or radial precursors in reactive plasma environments, and grow into larger particles and incredibly small crystallites \cite{Vladimirov2004}. The existing plasma wave spectra have been significantly modified when charged dust particles are introduced. Consideration of charged dust grains in plasma not only modifies the current plasma wave spectra but also provides a variety of unique innovative eigenmodes, such as dust-acoustic (DA) waves \cite{Rao1990}, dust-ion acoustic (DIA) \cite{Shukla2009}, dust lattice, etc.  Rao \emph{et al.} \cite{Rao1990} at first predicted the presence of this unique extremely low-phase-velocity (as compared to the thermal velocities of electrons and ions) DA waves, in which the dust mass provides the inertia and the electron and ion thermal pressures produce the restoring force. Afterward, many laboratory investigations have thoroughly proven Rao \emph{et al.}'s \cite{Rao1990} prediction \cite{Barkan1995,Merlino2004}.

Over the last two decades, several authors have investigated nonlinear DA waves in various dusty plasma environments, both theoretically \cite{Alinejad2010,Eliasson2004,Ma1997,El-Labany2003,Mamun2008} and experimentally \cite{Bandyopadhyay2008,Heinrich2009}. Usually, dust particles are thought to be massively negatively charged objects because of the accumulation of electrons from background plasma species \cite{Bharuthram1992}. However, there are a number of mechanisms (such as photoemission in the presence of ultraviolet light or thermionic emission from grains heated by radiation) by which a dust particle can acquire a positive charge and coexist with negatively charged dust particles, ions, and electrons in a variety of dusty plasmas (DP), such as the Earth's mesosphere \cite{Havnes1996,Gelinas1998,Mendis2004}, cometary tail \cite{Horanyi1996,Mamun2004}, Jupiter's surrounds \cite{Tsintikidis1996}, Jupiter's magnetosphere \cite{Horanyi1993}, etc., and laboratory equipment \cite{Khrapak2001,Fortov2003,Davletov2018}. As a consequence of the following three key mechanisms, the dust species get positively charged \cite{Chow1993,Rosenberg1995,Rosenberg1996,Samarian2001}.

  1. The collision of highly energetic plasma particles like electrons or ions results in the secondary emission of electrons from the surface of the dust grain \cite{Chow1993}.

  2. The severe radiative or thermal heating that induces the thermionic emission of electrons from the surface of the dust grain \cite{Rosenberg1995}.

  3. Electron photoemission from the dust grain surface is caused by a flow of high-energy photons \cite{Rosenberg1996}.

The existence of ions and electrons that are not in thermodynamic equilibrium is exposed by space plasma observations \cite{Asbridge1968,Feldman1983,Lundin1989,Franz1998} defining aspect of space plasma is the nonthermal or superthermal distribution functions that the electrons and ions follow. Such distribution functions are a well-known characteristic of the auroral zone, to be exact \cite{Hall1991}. It is becoming recognized that the nonthermal electron/ion distributions constitute a distinctive aspect of space plasmas. Nonthermal velocity distributions are typically simple to evaluate using the Cairns distribution function \cite{Cairns1995}. The study of nonthermal ions and electrons in dusty plasma is therefore becoming more and more relevant. The Cairns velocity distribution function, which may be represented identically in 1D normalized form as \cite{Cairns1995}:

\begin{eqnarray}
  &&\hspace*{-1.5cm}f(v) = \frac{1+\alpha (v^{2}- 2 \phi)^{2}}{(1+3\alpha)\sqrt{2 \pi}} \exp\left[-\frac{1}{2}(v^2 - 2\phi)\right]\,.
  \label{1}
\end{eqnarray}
Here, $ \phi$ is the electrostatic wave potential, while $\alpha$ is a parameter determining the number of fast (energetic) particles in the plasma system under study. Taibany and Sabry \cite{El-Taibany2005} studied the small amplitude dust-acoustic solitary waves and double layers with the help of the Zakharov-Kuznetsov equation. It is observed that due to the nonthermal ions both compressive and rarefactive solitary waves exist. Kian and Mahdieh \cite{Kian2022} has shown that both non-thermal distributed electron and ion significantly modify the properties of large amplitude dust-acoustic waves. Verheest \cite{Verheest2009} has considered both nonthermal electrons and ions to study the large amplitude dust-acoustic solitary waves and double layers in opposite polarity dusty plasma system. The effect of the Cairns nonthermal distribution of the ion species on the ion-acoustic solitary waves has been studied by Mamun and Mannan \cite{Mamun2021}.

The existence of positively charged dust species in electron-ion plasmas has attracted many plasma physicists to explore the new features of linear and nonlinear ion-acoustic waves. The presence of positively charged dust species plays a vital role in modifying the salient features of these kinds of waves \cite{Mamun2021a,Mamun2021,Mushinzimana2022,Tarofder2023}. Mamun \cite{Mamun2021a} found that the stationary positively charged dust species plays a vital role in the formation of ion-acoustic subsonic solitary waves in electron-ion-positively charged dust plasma medium. Mamun and Mannan \cite{Mamun2021} studied the effects of static positively charged dust species in a complex plasma medium. They observed that the presence of static positively charged dust species in complex plasma systems supports the ion-acoustic solitary waves and double layers. Mushinzimana \textit{et al.} \cite{Mushinzimana2022} studied the propagation of dust-ion-acoustic solitary waves and double layers in a dusty plasma with the presence of adiabatic positively charged dust grains. Recently, Susmita \textit{et al.} \cite{Tarofder2023} investigated the propagation of three-dimensional cylindrical dust-acoustic solitary waves in an unmagnetized dusty plasma environment. The latter is composed of positively charged adiabatic dust grains, nonthermal ions, and electrons. They have used the reductive perturbation method which is valid for small amplitude limits.

In this paper, we study the large or arbitrary amplitude dust-acoustic waves in a warm nonthermal plasma medium containing warm adiabatic positively charged dust species, Carins nonthermal distributed electrons and ions. To reduce our set of governing equations that describe our plasma system into an energy integral equation we use the well-known and widely used method so called the pseudo-potential approach. We then present the formation of solitary waves and double layers with the help of Sagdeev potential. We also present the formation of subsonic and supersonic solitary waves that depends on the critical Mach number. It is observed that the presence of warm adiabatic positively charged dust species significantly modifies the formation of potential wells and basic features (viz. amplitude, width, speed, etc.) of positive and negative solitary waves and double layers.

The manuscript is organized as follows. In Section \ref{2sec:Governing Equations}, we present the model equations that describe our plasma system under consideration. The formation and properties of solitary waves and double layers by using the pseudo-potential technique are discussed in Section \ref{3sec:DA Solitary waves}. Finally, the results of our present theoretical investigation are reported in Section \ref{4sec:Discussion}.

\section{Governing Equations}
\label{2sec:Governing Equations}
To study the nonlinear propagation of DA waves we consider a collisionless, unmagnetized nonthermal dusty plasma medium containing adiabatic positively charged dust species and nonthermal distributed electrons and ions. According to our present plasma system, the charge neutrality condition at equilibrium can be written as $n_{e0} = Z_{d}n_{d0} + n_{i0}$, where $n_{e0}$, $n_{d0}$, and $n_{i0}$ are the unperturbed number densities of the nonthermal electrons, PCD species, and nonthermal ions, respectively and $Z_d$ is the number of electrons residing onto the positive dust grain surface. The positively charged dust grain dynamics are governed by the following dimensionless equations:

\begin{eqnarray}
&&\hspace*{-1.5cm}\frac{\partial n_{d}}{\partial t}+\frac{\partial}{\partial x}{(n_{d}u_{d})} = 0 ,\label{2}\\
&&\hspace*{-1.5cm}\frac{\partial u_{d}}{\partial t}+ u_{d}\frac{\partial u_{d}}{\partial x} = - \frac{\partial \phi}{\partial x}-\frac{\sigma_{d}}{n_{d}}\frac{\partial {n_{d}}^{\gamma}}{\partial x}, \label{3}\\
&&\hspace*{-1.5cm}\frac{\partial^{2} \phi}{\partial x^{2}} = (1+\mu)n_{e} - \mu n_{i} -n_{d}. \label{4}
\end{eqnarray}
For convenience, we use here dimensionless equations by introducing the normalizing factors as follows:
$n_e$, $n_i$, and $n_d$ are normalized by their unperturbed number density $n_{e0}$, $n_{i0}$, and $n_{d0}$, respectively; dust fluid velocity $u_d$ is normalized by the DA speed $C_{d} = (Z_{d} k_{B} T_{i} /m_{d} )^{1/2}$ with $k_B$, $T_i$, and $m_d$ being the Boltzmann constant, ion temperature, and dust grain mass, respectively; the electrostatic wave potential $\phi$ is normalized by $k_{B}T_{i}/e$; the space $x$ and time $t$ variables are normalized by Debye radius $\lambda_d = (k_BT_i/4\pi n_{d0}Z_de^2)^{1/2}$ and dust plasma period $\omega_{pd}^{-1}=(m_d/4\pi n_{d0}Z_d^2e^2)^{1/2}$, respectively; the dust adiabatic index $\gamma = (2+N)/N$ is equal to 3 for one-dimensional cases in our present study, where $N$ is the number of degrees of freedom; $\sigma_d=T_d/Z_dT_i$ with $T_d$ being the dust temperature; and $\mu = n_{i0}/Z_dn_{d0}$.

Besides the warm adiabatic dust, both nonthermal cairns distributed electrons and ions in the dimensionless form as follows:
\begin{eqnarray}
  &&\hspace*{-2cm}n_{e}=(1-\sigma_{i}\beta\phi + \beta(\sigma_{i}\phi)^{2}) \exp(\sigma_{i}\phi)\,,  \label{5}\\
  &&\hspace*{-2cm}n_{i}=(1+\beta\phi+\beta{\phi}^{2}) \exp(-\phi)\,, \label{6}
\end{eqnarray}
where $\beta = 4\alpha/(1+3\alpha)$ is the nonthermal parameter and $\sigma_i = T_i/T_e$.
\section{DA solitary waves and Double layers}
\label{3sec:DA Solitary waves}
To analyze the fully nonlinear DA solitary waves and double layers by using the pseudo-potential method \cite{Sagdeev1966,Mannan2013,Bernstein1957,Mannan2020}, we assume a comoving frame where the nonlinear structure is stationary ($\partial/\partial t =0$) and all the variables to be undisturbed at $|\xi|\rightarrow \infty$. So, all the dependent variables in equations depend only on one variable $\xi = x - Mt$, where $M$ is the Mach number. Therefore, we write our dimensionless equations \eqref{2}-\eqref{4} in terms of new variable as

 \begin{eqnarray}
&&\hspace*{-0.5cm}M\frac{\partial n_{d}}{\partial \xi}- \frac{\partial}{\partial \xi} (n_{d}u_{d}) = 0, \label{7}\\
&&\hspace*{-0.5cm}M\frac{\partial u_{d}}{\partial \xi}- u_{d} \frac{\partial u_{d}}{\partial \xi} = \frac{\partial \phi}{\partial \xi} + \frac{\sigma_{d}}{n_{d}} \frac{\partial {n_{d}}^{3}}{\partial \xi}, \label{8}\\
&&\hspace*{-0.5cm}\frac{\partial^{2}\phi}{\partial \xi^{2}} = (1+\mu)n_{e} - \mu n_{i} -n_{d} \label{9}.
 \end{eqnarray}
Integrating equation \eqref{7} with respect to $\xi$ and using the appropriate boundary conditions, $n_d =1$, $u_d = 0$, $\phi(\xi) =0$, $d\phi/d\xi = 0$ at $|\xi|\rightarrow \infty$ we obtain the normalized dust fluid velocity
 \begin{equation}
 u_{d} = M - \frac{M}{n_{d}}  \label{10}.
 \end{equation}
By using the above-mentioned boundary conditions and \eqref{10} into equation (\ref{8}) one obtains
 \begin{equation}
 3\sigma_{d}n_{d}^4 - (M^2 + 3\sigma_{d} - 2\phi)n_{d}^2 + M^2 = 0\,. \label{11}
 \end{equation}
Thus the expression for normalized dust number density $n_{d}$ from \eqref{11} can be expressed as
 \begin{equation}
   n_{d} = \frac{1}{\sqrt{6\sigma_{d}}} \left(\psi - \sqrt{\psi^{2} - 12\sigma_{d}M^{2}}\right)^{1/2} \label{12},
 \end{equation}
where $\psi =( M^{2} + 3\sigma_{d} - 2\phi)$. By inserting equation \eqref{5}, \eqref{6}, and \eqref{12} into \eqref{9} and multiplying on both sides of the resulting equation by $d\phi/d\xi$ and integrating once with respect to $\xi$ with the above-mentioned boundary conditions, we obtain the following differential equation
\begin{equation}
\frac{1}{2}\left(\frac{d\phi}{d\xi}\right)^2 + V(\phi, M) = 0\,. \label{13}
\end{equation}
 This equation represents an energy integral of a pseudo-particle of unit mass, pseudo time $\xi$, pseudo-position $\phi$ and the pseudo-potential $ V(\phi, M)$ is defined by
\begin{align}
V(\phi,M) = C -\frac{e^{\sigma_{i}\phi}(1+\mu)}{\sigma_i}\left(1+3\beta-3\beta\sigma_i\phi+\beta\sigma_i^2\phi^2\right)
\nonumber\\
 - \frac{1}{3}\sqrt{\frac{2}{3\sigma_{d}}}\left(\psi + \frac{1}{2}\sqrt{\psi_1}\right)\left(\psi - \sqrt{\psi_1}\right)^{1/2}\nonumber\\
 - \mu e^{-\phi}\left(1 + 3\beta + 3\beta\phi + \beta \phi^{2}\right)\,,\label{sagdeev}
\end{align}
where $\psi_1 = \psi^{2} - 12\sigma_{d}M^{2}$ and the integration constant
\begin{equation}\label{14}
C = (1 + 3\beta) \left(\frac{1 + \mu}{\sigma_{i}} + \mu\right) + M^{2} + \sigma_{d}\,.
\end{equation}
Note that $C$ is determined in such a way that $V(\phi , M) = 0$ at $\phi = 0$. The Sagdeev potential $V(0, M) = 0$ is satisfied because of our choice $C$. Due to the equilibrium charge neutrality condition $V^{\prime}(0, M) = 0$ is also satisfied. Here the prime ($'$) denotes the derivative of $V(\phi ,M)$ w. r. to $\phi$. To obtain the solitary waves and double layers solution one has to choose the origin an unstable maximum, i.e. $V^{\prime\prime}(0, M) < 0$. The latter confirms the expression of Mach number $M_c$ that is the solution of $V^{\prime\prime}(0, M) = 0$ and $M_c$ is defined as
\begin{equation}\label{15}
M_{c} = \left(\frac{1}{(1 - \beta)(\mu + \sigma_{i} + \mu\sigma_{i})}+3\sigma_d\right)^{1/2}\,.
\end{equation}
It is worth mentioning that $V^{\prime\prime\prime}(0, M_{c} =0)$ allows the value of parameters that determine the sign changes in the polarity of solitary waves. Thus, we can write \cite{Cairns1995,Mamun1997}:
\begin{equation}\label{16}
 V^{\prime\prime\prime}(0, M_{c}) = \mu-(1+\mu)\sigma_i^2+(3+12\sigma_d\rho)\rho^2\,,
\end{equation}
where $\rho = (1-\beta)(\mu+\sigma_i+\mu\sigma_i)$. Note also that the conditions $V^{\prime}(\phi_{m}, M) > 0$ and $V^{\prime}(\phi_{m}, M) < 0$ determine the solitary wave with positive potential ($\phi > 0$) and negative potential ($\phi < 0$), respectively. At the same time, $V^{\prime}(\phi_{m}, M) = 0$ determines the double layers. Here $\phi_{m}$ denotes the amplitude of the solitary waves or double layers. Finally, it is concluded that the solitary waves and double layers exist if and only if $V^{\prime\prime}(0, M) < 0$, i.e. $M > M_{c}$, where $M_c$ is defined in \eqref{15}.

\begin{figure}[H]
\includegraphics[width=80mm]{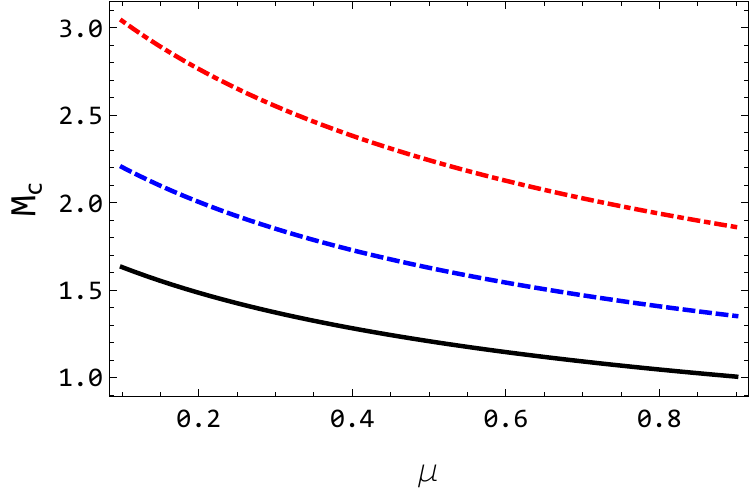}
\caption{The variation of $M_{c}$ with $\mu$ for $\sigma_{d} = 0.01$ , $\sigma_{i} = 0.6$, $\alpha = 0.2$ (solid curve), $\alpha = 0.4$ (dashed curve), and $\alpha = 0.6$ (dot-dashed curve).}
\label{Fig-1}
\end{figure}

\begin{figure}[H]
\includegraphics[width=80mm]{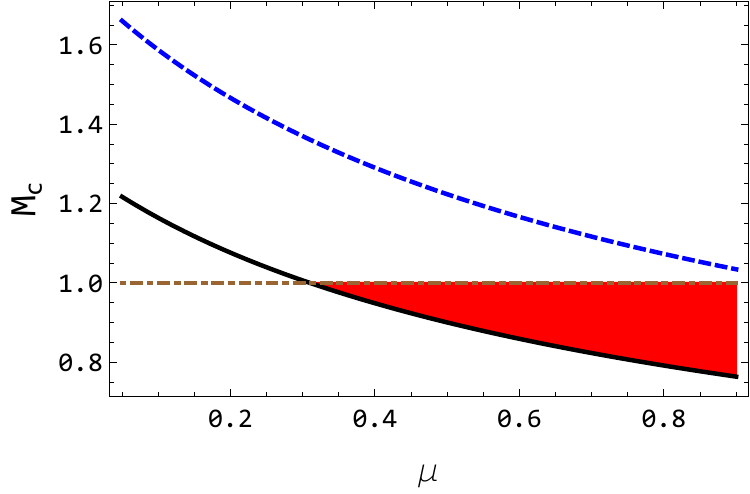}
\caption{The variation of $M_{c}$ with $\mu$ for $\sigma_{d} = 0.01$ , $\sigma_{i} = 0.9$, $\alpha = 0.1$ (solid curve), and $\alpha = 0.3$ (dashed curve).}
\label{Fig-1b}
\end{figure}

\begin{figure}[htb]
\includegraphics[width=80mm]{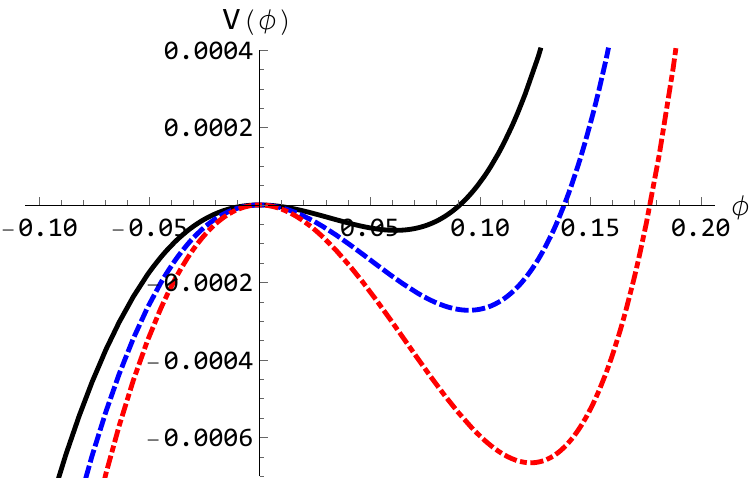}
\caption{The formation of the potential wells representing the subsonic SWs with $\phi>0$ for $\alpha = 0.1$ , $M = 0.99$ , $\sigma_{d} = 0.01$ , and $\sigma_{i} = 0.9$, $\mu = 0.4$ (solid curve), $\mu = 0.45$ (dashed curve), $\mu = 0.5$ (dot-dashed curve).}
\label{Fig-3}
\end{figure}

Figures \ref{Fig-1} and \ref{Fig-1b} visualize how the Mach number or phase speed of DA waves changes with different values of nonthermal parameter $\alpha$ and ion to positively charged dust number density ratio $\mu$. It is found that $M_c$ decreases as $\mu$ increases. Note that as we increase the values of $\sigma_d$, $M_c$ also increases but $M_c$ decreases with $\sigma_i$. It is worth mentioning that the effects of nonthermality $\alpha$ and $\mu$ determine the formation of subsonic and supersonic DA waves. The conditions $M_c<M<1$ and $1<M_c<M$ determine the subsonic and supersonic DA waves, respectively. It is observed that the region of formation of subsonic DA waves becomes broadened when the nonthermal parameter $\alpha$ decreases. On the contrary, the possibility of the formation of supersonic DA waves becomes very high with the increasing values of ion number density and nonthermality. The red color shaded area (as shown in figure \ref{Fig-1b}) represents the region of the formation of subsonic DA waves.
\begin{figure}[H]
\includegraphics[width=80mm]{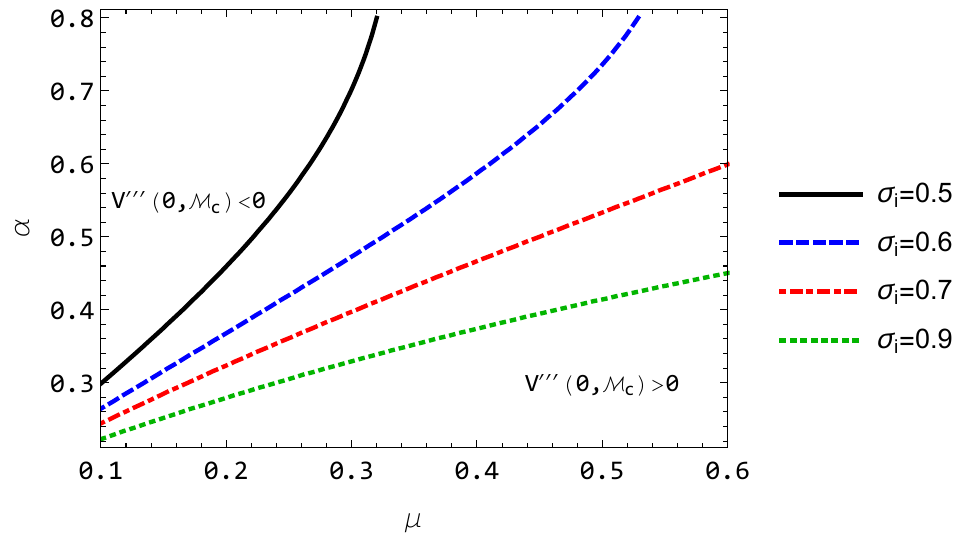}
\caption{The contour plot of $V^{\prime\prime\prime}(0, M_{c}) = 0$ is plotted against $\mu$ and $\alpha$ for $\sigma_d=0.01$, $\sigma_i = 0.5$ (solid curve), $\sigma_i = 0.6$ (dashed curve), $\sigma_i = 0.6$ (dot-dashed curve), and $\sigma_i = 0.9$ (dotted curve).}
\label{Fig-2}
\end{figure}
The expression in \eqref{16} determines the polarity of subsonic and supersonic DA solitary waves and double layers. Figure \ref{Fig-3} shows how the regions of $V^{\prime\prime\prime}(0, M_{c}) > 0$, $V^{\prime\prime\prime}(0, M_{c}) = 0$, and $V^{\prime\prime\prime}(0, M_{c}) < 0$ change with $\mu$ and $\alpha$ for different values of $\sigma_i$. It is mentioned that $V^{\prime\prime\prime}(0, M_{c}) > 0$ ($V^{\prime\prime\prime}(0, M_{c}) > 0$) determine the positive (negative) solitary waves and double layers potentials. Increasing the value of $\sigma_i$ reduces the formation of positive solitary waves and double layers regimes.
\begin{figure}[htb]
\includegraphics[width=80mm]{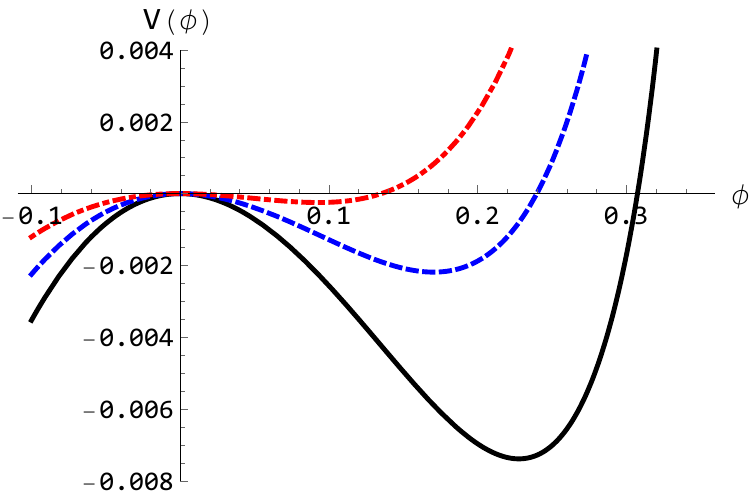}
\caption{The formation of the potential wells representing the subsonic SWs with $\phi>0$ for $\mu = 0.8$ , $M = 0.99$ , $\sigma_{d} = 0.01$ , and $\sigma_{i} = 0.9$ , $\alpha = 0.1$ (solid curve), $\alpha = 0.15$ (dashed curve), $\alpha = 0.2$ (dot-dashed curve).}
\label{Fig-4}
\end{figure}
\begin{figure}[htb]
\includegraphics[width=80mm]{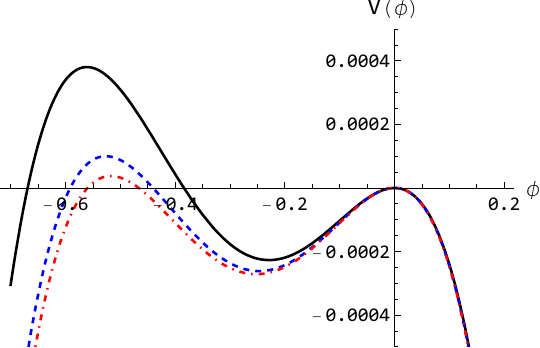}
\caption{The formation of the potential wells representing the supersonic SWs with $\phi<0$ for $\sigma_{d} = 0.01$, $\sigma_{i} = 0.7$, $\alpha =0.6$, $M = 3.31$, $\mu = 0.101$ (solid curve), $\mu = 0.105$ (dashed curve), and $\mu = 0.106$ (dot-dashed curve).}
\label{Fig-5}
\end{figure}
\begin{figure}[htb]
\includegraphics[width=80mm]{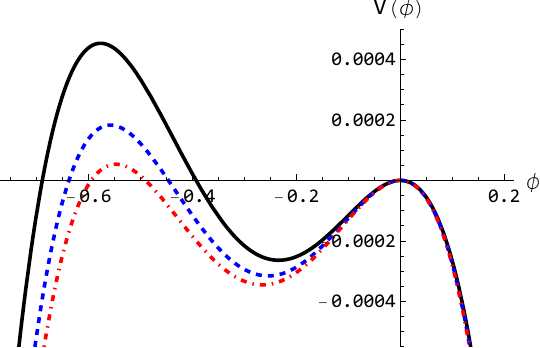}
\caption{The formation of the potential wells representing the supersonic SWs with $\phi<0$ for $\sigma_{d} = 0.01$, $\sigma_{i} = 0.72$, $\mu = 0.1$, $M = 3.5$, $\alpha =0.62$ (solid curve), $\alpha =0.616$ (dashed curve), and $\alpha =0.614$ (dot-dashed curve).}
\label{Fig-6}
\end{figure}
\begin{figure}[htb]
\includegraphics[width=80mm]{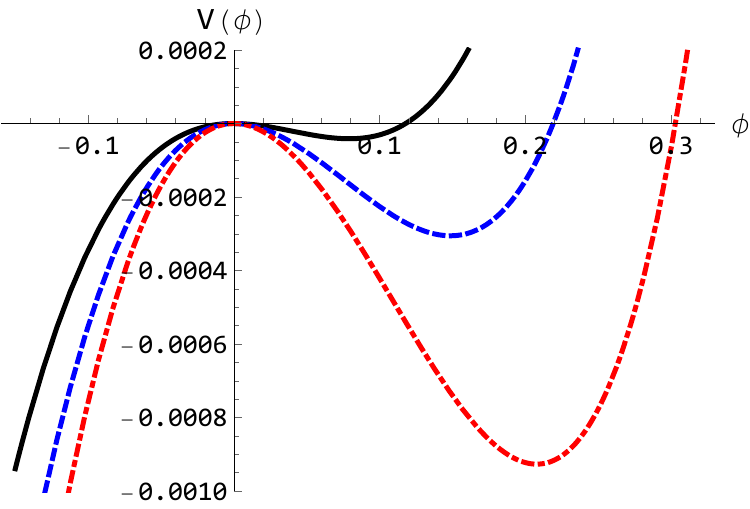}
\caption{The formation of the potential wells representing the supersonic SWs with $\phi>0$ for $\sigma_{d} = 0.01$ , $\alpha = 0.2$ , $\sigma_{i} = 0.6$, $M = 1.323$, $\mu =0.4$  (solid curve), $\mu= 0.45$ (dashed curve), $\mu = 0.5$ (dot-dashed curve).}
\label{Fig-7}
\end{figure}
\begin{figure}[htb]
\includegraphics[width=80mm]{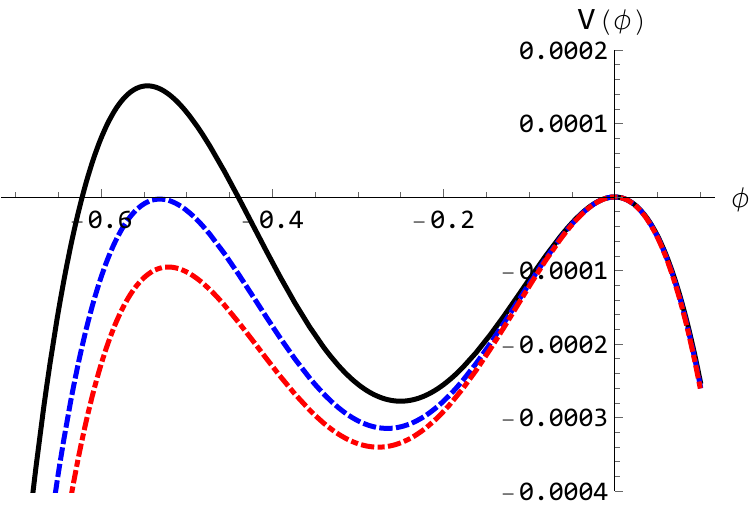}
\caption{The formation of the pseudo-potential wells with $\phi<0$ representing the supersonic double layers for $\mu = 0.1$, $\sigma_{i} = 0.7$, $\sigma_d=0.01$, $\alpha =0.6$, $M =3.35$  (solid curve), $M= 3.372$ (dashed curve), $M = 3.386$ (dot-dashed curve).}
\label{Fig-8}
\end{figure}
The Sagdeev potential $V(\phi)$ (represented in \eqref{sagdeev}) wells are plotted against the pseudo-position $\phi$. The formation of Sagdeev potential wells corresponding to the positive ($\phi>0$) and negative ($\phi<0$) solitary waves and double layers are shown in figures \ref{Fig-3} -\ref{Fig-8}. The distance between the intercept on the positive and negative $\phi$-axes and origin is the amplitude $\phi_m$ of solitary waves or double layers. The width is defined by $(\phi_{m}/\sqrt{|V_{m}|})$, where $|V_{m}|$ is the highest possible value of $V(\phi)$ in the pseudo-potential wells. The Sagdeev potential wells in the positive $\phi$-axis correspond to the formation of the positive subsonic solitary waves (as displayed in figures \ref{Fig-3} and \ref{Fig-4}). The amplitude (width) of the positive subsonic solitary waves increases (decreases) with the increasing value of $\mu$. But the effect of nonthermal parameter $\alpha$ reduces (enhances) the amplitude (width) of subsonic solitary waves with $\phi>0$.
The positive and negative supersonic solitary waves correspond to the formation of the pseudo-potential wells with positive and negative $\phi$-axes (as displayed in figures \ref{Fig-5}-\ref{Fig-7}). The effects of nonthermal parameter $\alpha$ and $\mu$ on positive and negative supersonic solitary waves are found to be similar in figures \ref{Fig-3} and \ref{Fig-4}. The amplitude (width) of both subsonic and supersonic solitary waves decrease ( increase) with $\sigma_d$. For $M>M_c$, the negative double layer is formed with negative potential $\phi<0$ only (as displayed in figure \ref{Fig-8}). Figure \ref{Fig-8} provides that solitary wave solution is formed at Mach number $ M = 3.35$ (solid curve), double layer solution is found at $M = 3.372$ (dashed curve), and no solitary wave structure exists at M = 3.386 (dot-dashed curve). It is important to mention that no subsonic solitary waves with negative potential $\phi<0$ are observed for our choice of plasma parameters.

\section{Discussion}
\label{4sec:Discussion}
We have considered a more general and realistic nonthermal dusty plasma medium consisting of nonthermal distributed electrons and ions and warm adiabatic positively charged dust species. We have reduced the model equations describing our present plasma system into a well-known energy integral equation by employing the pseudo-potential approach. The formation and properties of arbitrary or large amplitude dust-acoustic solitary waves and double layers are investigated with the help of Sagdeev-potential. Theoretical and numerical results obtained from our present plasma system are summarized as:
\begin{enumerate}
\item The phase speed of dust-acoustic waves is greatly influenced by the effect of nonthermal electrons and ions. Due to the presence of a more fast population of electrons and ions, the phase speed of dust-acoustic waves increases. The presence of a positively charged dust particle also increases the phase speed of DA waves. On the contrary, the phase speed of DA waves reduces as we increase the ion number density. Because of adiabatic warm positively charged dust species, the Mach number increases, but increasing the ion temperature reduces the value of the Mach number. Depending on the value of the phase speed of DA waves, the formation of subsonic and supersonic DA waves is discussed.

\item The parametric regimes of how the polarity of DA solitary waves and double layers changes with plasma parameters are shown.

\item The pseudo-potential wells are formed in both positive and negative $\phi$-axes. The properties of the height and thickness in the potential well formed in both positive and negative $\phi$-axes are discussed. The solitary waves and double layers solution are found in the values of Mach number that are around its critical value ($M_c$), i.e. $M>M_c$. Depending on the value of Mach number both subsonic and supersonic DA solitary waves are observed. It is seen that the salient features (the amplitude, the width, the speed, etc.) of DA solitary waves are significantly modified the plasma parameters. The positive dust temperature reduces the amplitude of solitary waves but increases the width of solitary waves. Increasing the ion number density causes to increase (decrease) the amplitude (width) of the solitary waves. On the other hand, the amplitude (width) of the solitary waves decreases (increases) with the nonthermal parameter.

\item For fixed plasma parameters, the solitary wave, the double layer, and no solitary wave solution are found with different values of Mach number.

\item No negative subsonic solitary structures and positive double layers are observed in our present plasma system.

\end{enumerate}
Finally, it is concluded that the results of our theoretical and numerical investigations will help in understanding the fundamental characteristics of DA supersonic and subsonic solitary waves and double layers that are found in space environments, such as the Earth's mesosphere or ionosphere \cite{Havnes1996,Gelinas1998,Mendis2004}, cometary tails \cite{Horanyi1996}, Jupiter's surroundings \cite{Tsintikidis1996} and magnetosphere \cite{Horanyi1993}, as well as laboratory devices \cite{Khrapak2001,Fortov2003,Davletov2018}.
\section*{Declarations}
\noindent \textbf{Disclosure of potential conflict of interest:} The authors declare that there is no conflict of interest.\\
\textbf{Funding:} This study was not supported by any funding.\\
\textbf{Data availability:} The data that support the findings of this study are available
within the article.

\end{document}